\documentclass[prd,aps,a4paper,twocolumn,eqsecnum,nofootinbib,floatfix]{revtex4}  

\newif\ifusesec
\usesectrue  
   
\usepackage{graphicx} 
\usepackage{amsmath,amsfonts,amssymb}

\newcommand{\beq}{\begin{equation}}
\newcommand{\eeq}{\end{equation}}
\newcommand{\bea}{\begin{eqnarray}}
\newcommand{\eea}{\end{eqnarray}}

\begin{document}

\title{Higher-order tail contributions to the energy and angular momentum fluxes in a two-body scattering process}

\author{Donato Bini$^{1,2}$, Andrea Geralico$^1$}
  \affiliation{
$^1$Istituto per le Applicazioni del Calcolo ``M. Picone,'' CNR, I-00185 Rome, Italy\\
$^2$INFN, Sezione di Roma Tre, I-00146 Rome, Italy\\
}

\date{\today}

\begin{abstract}
The need for more and more accurate gravitational wave templates requires taking into account all possible contributions to the emission of gravitational radiation from a binary system.
Therefore, working within a multipolar-post-Minkowskian framework to describe the gravitational wave field in terms of the source multipole moments, the dominant instantaneous effects should be supplemented by hereditary contributions arising from nonlinear interactions between the multipoles.
The latter effects include tails and memories, and are described in terms of integrals depending on the past history of the source.
We compute higher-order tail (i.e., tail-of-tail, tail-squared and memory)  contributions to both energy and angular momentum fluxes and their averaged values along hyperboliclike orbits at the leading post-Newtonian approximation, using harmonic coordinates and working in the Fourier domain. Due to the increasing level of accuracy recently achieved in the determination of the scattering angle in a two-body system by several complementary approaches, the knowledge of these terms will provide useful information to compare results from different formalisms.
\end{abstract}

\maketitle

\section{Introduction}
 
Tail effects in a two-body interaction are generated in the wave zone far from the system, where the latter can be described as a single object endowed with multipoles.
The gravitational wave generation formalism developed by Blanchet and Damour~\cite{Blanchet:1985sp,Blanchet:1989ki,Damour:1990ji,Blanchet:1998in,Poujade:2001ie} combines a multipolar-post-Minkowskian (MPM) expansion in the far zone with a post-Newtonian (PN) expansion in the near zone to relate the gravitational radiation emitted by the binary system to the post-Newtonian expansion in the near zone, where the two constituents of the binary can be resolved as individual sources.  
A matching procedure in the overlapping region where both expansions are valid allows for expressing the radiative moments as non-linear functionals of two infinite
sets of time-varying source multipole moments.
The latter moments mix with each other as the waves propagate, so that the relation between radiative and source moments include many nonlinear interactions, which are called hereditary effects \cite{Blanchet:1987wq,Blanchet:1992br,Blanchet:1993ec,Blanchet:1994ez,Rieth:1997mk,Blanchet:1997jj}, depending on the full past history of the source.

Starting from the 4PN level of accuracy, the Hamiltonian governing the conservative two-body dynamics acquires a nonlocal part summarizing several such hereditary effects, including tails, tails-of-tails, tail-squared, memory etc. 
The dominant tails are due to the quadratic nonlinear interaction between higher-order multipole moments and the mass monopole, namely the total Arnowitt-Deser-Misner (ADM) mass. The nonlinear memory effect also arises at the quadratic level due to the interaction between two quadrupole moments.
The tail-of-tail and tail-squared contributions are cubic nonlinear effects caused by the interaction between the tail itself and the ADM mass and the self-interaction of the tail itself, respectively \cite{Blanchet:1997jj}. 

Tail-transported nonlocal dynamical correlations lead to a nonlocal action, so that the instantaneous interaction terms in the Hamiltonian are complemented by a (time-symmetric) nonlocal-in-time interaction \cite{Damour:2014jta,Jaranowski:2015lha,Damour:2016abl}.
The nonlocal Hamiltonian has been recently determined up to 6PN by using a time-split version of the gravitational-wave energy flux, including both first-order (4+5+6PN) and second-order (5.5PN) tail effects \cite{Bini:2019nra,Bini:2020wpo,Bini:2020nsb,Bini:2020hmy}.
The formalism developed there has allowed to compute both local and nonlocal parts of the conservative scattering angle up to the seventh order in $G$ by using a combined PM-PN expansion \cite{Bini:2020rzn}.
However, radiation-reaction effects as well as tail effects also enter the problem starting at $O(G^3)$ and $O(G^4)$, respectively.
This fact has both conservative and dissipative aspects, as discussed in Ref. \cite{Bini:2021gat}. 
Having already taken into account the time-symmetric aspect of tail interaction in the nonlocal contribution to the conservative dynamics, the only tail-related effect to be added to radiation-reaction is the time-antisymmetric one, as explicitly shown in Ref. \cite{Damour:2014jta} at 4PN level.

In order to evaluate the radiation-reaction contributions to the scattering angle one needs the radiative losses of energy, angular momentum and linear momentum.
While there exists a rich literature for the gravitational wave tails in the case of coalescing black holes \cite{Arun:2007rg,Arun:2009mc,Blanchet:2011zv,Marsat:2013caa,Blanchet:2017rcn,Marchand:2016vox}, the companion situation of black holes undergoing a scattering process is less studied. 
We have already computed in Ref. \cite{Bini:2021gat} the (integrated) leading-order tail contributions to the loss of energy, angular momentum and linear momentum along hyperboliclike orbits, limiting to the leading PN term for each of them.
We will refer to these terms as \lq\lq past tails," to be distinguished from the time-symmetric tails entering the nonlocal part of the Hamiltonian.
In the present paper we evaluate the orbital average of higher-order energy and angular momentum past tails (tail-of-tail and tail-squared) as well as the corresponding time-symmetric tails.

The paper is organized as follows.
In Sec. II we recall the main definitions of the various past tail integrals computed in this work.
These integrals are more conveniently computed in the frequency domain.
Section III provides all necessary information to get the final expressions for the hereditary contributions to the orbital-averaged energy and angular momentum fluxes, using a Fourier decomposition of the multipole moments.
The explicit evaluation for hyperboliclike motion is done in Sec. IV, where the results are expressed as an expansion in the large angular momentum parameter. 
Time-symmetric tails are instead computed in Sec. V, using a slight extension of the general formulas introduced in Sect. III.
Finally, in Sec. VI we summarize our results and discuss their relevance for recent developments in the relativistic two-body scattering problem.

We will denote the masses of the two bodies as $m_1$ and $m_2$ with $m_2>m_1$, and define the symmetric mass ratio $\nu={\mu}/{M}$ as the ratio of 
the reduced mass $\mu\equiv m_1 m_2/(m_1+m_2)$ to the total mass $M=m_1+m_2$, as standard.
We will use the following dimensionless energy and angular momentum parameters 
\beq \label{defbarE}
\bar E \equiv \frac{E_{\rm tot}-Mc^2}{\mu c^2}\,,
\eeq
and 
\beq \label{defj2}
j\equiv \frac{c J}{G m_1 m_2}=\frac{c J}{G M \mu}\,,
\eeq
where $E_{\rm tot}$ and $J$ are the total center-of-mass energy and angular momentum of the binary system, respectively.

\section{Energy and angular momentum tail integrals}

The total contribution to the energy and angular momentum fluxes
\beq
\mathcal{F}\equiv\left(\frac{d E}{dt}\right)^{\rm GW}\,, \qquad 
\mathcal{G}_i\equiv\left(\frac{d J_i}{dt}\right)^{\rm GW}\,,
\eeq
can be split as the sum of instantaneous and hereditary terms.
The latter can be further decomposed as tail, tail-of-tail, tail-squared and higher non-linear interaction terms.
We will consider below quadratic and cubic-in-$G$ interactions only at their leading PN level of accuracy.

The hereditary part of the gravitational wave energy flux reads
\beq
\mathcal{F}_\mathrm{hered}(t)=\mathcal{F}_{\rm tail}(t)+\mathcal{F}_\mathrm{tail(tail)}(t)+\mathcal{F}_\mathrm{(tail)^2}(t)\,,
\eeq
where the quadratic and cubic tails (according to the terminology introduced in Ref. \cite{Blanchet:1997jj}) are given by
\begin{widetext}
\bea
\label{Ftail}
\mathcal{F}_\mathrm{tail}(t) &=&  
\frac{G^2\,\mathcal{M}}{c^8}\,\Biggl\{\frac{4}{5}\,I_{ij}^{(3)}(t)\,\int_{-\infty}^td\tau\,I^{(5)}_{ij}(\tau)\left[\ln\left(\frac{t-\tau}{2\tau_0}\right)+\frac{11}{12}\right]
+O\left(\frac{1}{c^2}\right)\,\Biggr\}
\,,
\eea
and
\bea
\label{Ftailtail} 
\mathcal{F}_\mathrm{tail(tail)}(t) &=&
\frac{4}{5}\,\frac{G^3\mathcal{M}^2}{c^{11}}\,I_{ij}^{(3)}(t)\,\int_{-\infty}^td\tau\,I^{(6)}_{ij}(\tau)\left[
\ln^2\left(\frac{t-\tau}{2\tau_0}\right)+\frac{57}{70}\ln\left(\frac{t-\tau}{2\tau_0}\right)+\frac{124627}{44100}\right]
\,,\nonumber\\
\mathcal{F}_\mathrm{(tail)^2}(t) &=& 
\frac{4}{5}\,\frac{G^3\mathcal{M}^2}{c^{11}}\,\left(\int_{-\infty}^td\tau\,I^{(5)}_{ij}(\tau) \left[\ln\left(\frac{t-\tau}{2\tau_0}\right)+\frac{11}{12}\right]\right)^2
\,,
\eea
respectively.
Here $\mathcal{M}$ denotes the total ADM mass of the system (which can be set equal to $M$ at the leading order in the PN expansion) and $\tau_0=cr_0$, with $r_0$ a constant length scale entering the relation between the retarded time in radiative coordinates and the corresponding retarded time in harmonic coordinates.
The quadratic term \eqref{Ftail} is the dominant tail at order 1.5PN, while the two cubic-order tails \eqref{Ftailtail} are both at 3PN order.

Similarly, the hereditary part of the angular momentum flux is decomposed as
\beq
{\cal G}_i^{\rm hered}(t)={\cal G}_i^{\rm tail}(t)+{\cal G}_i^{\rm tail(tail)}(t)+{\cal G}_i^{\rm (tail)^2}(t)+{\cal G}_i^{\rm memory}(t)\,,
\eeq
where 
\begin{eqnarray}
\label{Gitail}
{\cal G}_i^{\rm tail}(t)&=&\frac{G^2 \mathcal{M}}{c^8}\,\epsilon_{iab}\,\Biggl\{
\frac{4}{5}\,I_{aj}^{{(2)}}(t)\int_{-\infty}^{t} d\tau\left[{\ln}\left(\frac{t-\tau}{2 \tau_0}\right)+\frac{11}{12}\right]I_{bj}^{{(5)}}(\tau)\nonumber\\
&&
+\frac{4}{5}\,I_{bj}^{{(3)}}(t)\int_{-\infty}^{t} d\tau\left[{\ln}\left(\frac{t-\tau}{2 \tau_0}\right)+\frac{11}{12}\right]I_{aj}^{{(4)}}(\tau)
+O\left(\frac{1}{c^2}\right)
\Biggr\}
\,,
\end{eqnarray}
starting at 1.5PN order, and
\begin{eqnarray}
\label{Gitailtail}
{\cal G}_i^{\rm tail(tail)}(t)&=&
\frac{4}{5}\frac{G^3\mathcal{M}^2}{c^{11}}\,\epsilon_{iab}\,\Biggl\{
I_{aj}^{{(2)}}(t)\int_{-\infty}^{t}d \tau\left[\ln^2 \left(\frac{t-\tau}{2 \tau_0}\right) +\frac{57}{70} \ln\left(\frac{t-\tau}{2 \tau_0}\right)+\frac{124627}{44100}\right]I_{bj}^{{(6)}}(\tau)\nonumber\\
&&
+I_{bj}^{{(3)}}(t)\int_{-\infty}^{t}d \tau\left[\ln^2 \left(\frac{t-\tau}{2 \tau_0}\right) +\frac{57}{70} \ln\left(\frac{t-\tau}{2 \tau_0}\right)+\frac{124627}{44100}\right]I_{aj}^{{(5)}}(\tau) 
\Biggr\}
\,,\nonumber\\ 
{\cal G}_i^{\rm(tail)^2}(t)&=& 
\frac{8}{5}\frac{G^3\mathcal{M}^2}{c^{11}}\epsilon_{iab}
\left(\int_{-\infty}^{t} d \tau\left[\ln\left(\frac{t-\tau}{2 \tau_0}\right)+\frac{11}{12}\right]I_{aj}^{{(4)}}(\tau)\right)
\left(\int_{-\infty}^{t} d \tau\left[\ln\left(\frac{t-\tau}{2 \tau_0}\right)+\frac{11}{12}\right] I_{bj}^{{(5)}}(\tau)\right)
\,.
\end{eqnarray}
which are both 3PN order.
\end{widetext}
At the quadratic-in-$G$ order one also have the following nonlinear memory integral
\beq
\label{Gimemory}
{\cal G}_i^{\rm memory}(t)=
\frac{4}{35}\frac{G^2}{c^{10}}\,\epsilon_{iab}\,I_{aj}^{{(3)}}(t)\int_{-\infty}^{t} d \tau I_{c b}^{{(3)}}(\tau)\, I_{j c}^{{(3)}}(\tau)
\,,
\eeq
which is 2.5PN order.
Notice that there is no memory contribution in the case of the energy flux, because the memory integral is time differentiated and therefore becomes instantaneous, as discussed in Ref. \cite{Arun:2007rg}, and then taken into account in the instantaneous part.

It is convenient to introduce the following notation
\beq
\label{typical_int}
{\mathcal T}_{\ln^m{}}[X^{(n)}_L;C_{X_L}](t)=\int_{-\infty}^t d\tau X^{(n)}_L(\tau)\ln^m \left(\frac{t-\tau}{C_{X_L}}\right)\,,
\eeq
where $X^{(n)}_L$ denotes a generic multipolar moment with $L$ (either electric-type or magnetic-type) tensorial indices and differentiated $n$ times with respect to time, and $C_{X_L}$ is a constant which depends on the multipolar moment considered.
The integral ${\mathcal T}_{\ln^m{}}(X^{(n)}_L)$ thus represents the $m$-type past-tail associated with the history of $X_L^{(n)}$, from past infinity to the present time. 
For the purpose of the present work we only need $m=1,2$. 

The energy and angular momentum past tails (Eqs. \eqref{Ftail}--\eqref{Ftailtail} and \eqref{Gitail}--\eqref{Gitailtail}, respectively) can then be written as
\bea\label{Ftail_new}
\mathcal{F}_\mathrm{tail}(t) &=& \frac{4}{5}\, \frac{G^2\,{\mathcal M}}{c^8}\,I_{ij}^{(3)}(t) {\mathcal T}_{\ln{}}[I^{(5)}_{ij};C_{I_2}](t)
\,,\nonumber\\
\mathcal{F}_{\rm tail(tail)}(t)&=&\frac{4}{5}\,\frac{G^3\mathcal{M}^2}{c^{11}}\,I_{ij}^{(3)}(t)\,\left[
{\mathcal T}_{\ln^2{}}[I^{(6)}_{ij};C_{I_2}](t)\right.\nonumber\\
&&\left.
-\frac{107}{105}{\mathcal T}_{\ln{}}[I^{(6)}_{ij};\widetilde C_{I_2}](t)\right]
\,,\nonumber\\
\mathcal{F}_\mathrm{(tail)^2}(t) &=& \frac{4}{5}\,\frac{G^3{\mathcal M}^2}{c^{11}}\left({\mathcal T}_{\ln{}}[I^{(5)}_{ij};C_{I_2}](t) \right)^2
\,,\nonumber\\
\eea
and
\begin{widetext}
\bea
\label{Gitail_new}
{\mathcal G}_i^{\rm tail}(t) &=& \frac{4}{5}\,\frac{G^2 \mathcal{M}}{c^8}\epsilon_{iab}\left[
I^{(2)}_{aj}(t) {\mathcal T}_{\ln{}}[I^{(5)}_{bj};C_{I_2}](t)+I^{(3)}_{bj}(t){\mathcal T}_{\ln{}}[I^{(4)}_{aj};C_{I_2}](t) 
\right]
\,,\nonumber\\
{\mathcal G}_i^{\rm tail(tail)}(t) &=& \frac{4}{5}\,\frac{G^3\mathcal{M}^2}{c^{11}} \epsilon_{iab}\left\{
I^{(2)}_{aj}(t) \left[{\mathcal T}_{\ln^2{}}[I^{(6)}_{bj};C_{I_2}](t)-\frac{107}{105}{\mathcal T}_{\ln{}}[I^{(6)}_{bj};\widetilde C_{I_2}](t)\right]\right.\nonumber\\
&&\left.
+I^{(3)}_{bj}(t) \left[{\mathcal T}_{\ln^2{}}[I^{(5)}_{aj};C_{I_2}](t)-\frac{107}{105}{\mathcal T}_{\ln{}}[I^{(5)}_{aj};\widetilde C_{I_2}](t)\right]
\right\}
\,,\nonumber\\
{\mathcal G}_i^{\rm (tail)^2}(t) &=&\frac{8}{5}\,\,\frac{G^3\mathcal{M}^2}{c^{11}}\epsilon_{iab}{\mathcal T}_{\ln}[I^{(4)}_{aj};C_{I_2}](t)\,{\mathcal T}_{\ln}[I^{(5)}_{bj};C_{I_2}](t)
\,,
\eea
\end{widetext}
respectively, with
\beq
C_{I_2}=2\tau_0e^{-\frac{11}{12}}\,, \quad
\widetilde C_{I_2}=C_{I_2}e^{\frac{515063}{179760}}
\approx17.55C_{I_2}\,.
\eeq

In order to compute the above tail integrals at their leading PN approximation one only needs the quadrupole moment and its time derivatives evaluated at the Newtonian level.

We will evaluate below the leading order contribution to the orbital average of the tail integrals
\bea
\label{LOtailsaver}
(\Delta E)_\mathrm{X}&=&\int_{-\infty}^{\infty}dt\, \mathcal{F}_\mathrm{X} (t)
\,,\nonumber\\
(\Delta J_i)_\mathrm{X}&=&\int_{-\infty}^{\infty}dt\, \mathcal{G}_i^\mathrm{X} (t)
\,,
\eea
with $X=$ [tail, tail(tail), (tail)$^2$, memory], along hyperboliclike orbits.

\section{Computing the tail integrals in the Fourier domain}

Each of the integrals above is conveniently computed in the Fourier domain. 
Inserting in Eq. \eqref{typical_int} the Fourier expansion of $X_{L}(\tau)$, i.e., 
\beq
X_{L}(\tau)=\int_{-\infty}^\infty \frac{d\omega}{2\pi} \, e^{-i\omega  \tau}  \hat X_{L}(\omega) \,,
\eeq
and changing the integration variable as $t-\tau=\xi$ yield
\bea
\label{typical_int_n}
&&{\mathcal T}_{\ln^m}[X^{(n)}_L;C_{X_L}](t)
=\int_{0}^\infty d\xi X^{(n)}_L(t-\xi)\ln^m \left(\frac{\xi}{C_{X_L}}\right)\nonumber\\
&&\qquad\qquad
=\int_{-\infty}^\infty \frac{d\omega}{2\pi}e^{-i\omega  t  }(-i\omega)^n\hat X_L(\omega) A_m (\omega, C_{X_L})\,,\nonumber\\
\eea
with
\beq
A_m (\omega, C_{X_L})=\int_{0}^\infty d\xi  e^{ i\omega  \xi  } \ln^m \left(\frac{\xi}{C_{X_L}}\right)\,.
\eeq
For $m=1,2$ we have the relations (see Eqs. (4.7) and (4.13) of Ref. \cite{Arun:2007rg})
\bea
A_1 (\omega, C_{X_L})&=&-\frac{\pi}{2|\omega|}-\frac{i}{|\omega|}{\rm sgn}(\omega)\ln (C_{X_L}|\omega| e^\gamma)
\,, \nonumber\\
A_2 (\omega, C_{X_L})&=& \frac{\pi}{|\omega|}\ln (C_{X_L}|\omega| e^\gamma)\nonumber\\
&+&
\frac{i}{|\omega|}{\rm sgn}(\omega)\left[\ln^2 (C_{X_L}|\omega| e^\gamma)-\frac{\pi^2}{12}  \right]
\,,\nonumber\\
\eea
with the properties
\bea
\label{prop_of_A_n}
A_1(\omega, C_{X_L})+A_1(-\omega, C_{X_L})&=&-\frac{\pi}{|\omega|}\,,\nonumber\\
A_1(\omega, C_{X_L})-A_1(-\omega, C_{X_L})&=& -2\frac{i}{\omega}\ln (C_{X_L}|\omega| e^\gamma)\,, \nonumber\\
A_1(-\omega, C_{X_L})A_1(\omega, C_{X_L})&=&\frac{1}{\omega^2}\left(\frac{\pi^2}{4}\right.\nonumber\\
&+&\left.
\ln^2(C_{X_L}|\omega|e^\gamma)\right)\,,\nonumber\\
\eea
and
\beq
A_2(\omega, C_{X_L})=-i\omega \left(A_1(\omega, C_{X_L})^2-\frac{\pi^2}{6\omega^2}\right)\,.
\eeq

Taking the orbital averages \eqref{LOtailsaver} leads to integrals of the type
\begin{widetext}
\bea
\label{basic_int}
F_m[Y^{(p)}_M,X^{(n)}_L; C_{X_L}]&=&\int_{-\infty}^\infty dt\, Y^{(p)}_M(t)\, {\mathcal T}_{\ln^m}[X^{(n)}_L; C_{X_L}](t)\nonumber\\
&=& \int_{-\infty}^\infty dt\, \int_{-\infty}^\infty \frac{d\omega'}{2\pi}e^{-i\omega' t} (-i\omega')^p \hat Y_M(\omega') 
\,\int_{-\infty}^\infty  \frac{d\omega}{2\pi}e^{-i\omega  t  }(-i\omega)^n\hat X_L(\omega) A_m (\omega, C_{X_L})\nonumber\\
&=& (-1)^n  \int_{-\infty}^\infty  \frac{d\omega}{2\pi} (i\omega)^{n+p} \hat Y_M(-\omega) \hat X_L(\omega) A_m (\omega, C_{X_L})\nonumber\\
&=& \int_{0}^\infty  \frac{d\omega}{2\pi} (i\omega)^{n+p}[(-1)^n \hat Y_M(-\omega) \hat X_L(\omega) A_m (\omega, C_{X_L})+(-1)^p \hat Y_M(\omega) \hat X_L(-\omega) A_m (-\omega, C_{X_L})
]\,,\nonumber\\
\eea
which in the special case $Y=X$ and $L=M$ becomes
\bea
\label{fund_Y_eq_X}
F_m[X^{(p)}_L,X^{(n)}_L; C_{X_L}]
&=& \int_{0}^\infty  \frac{d\omega}{2\pi} (i\omega)^{n+p} \hat X_L(-\omega)\hat X_L(\omega)[(-1)^n  A_m (\omega, C_{X_L})+(-1)^p A_m (-\omega, C_{X_L})
]\,.
\eea
Using this result the energy and angular momentum past tails \eqref{Ftail_new}--\eqref{Gitail_new} become
\bea
\label{final_en_int}
(\Delta E)_{\rm tail} &=&  
\frac{2}{5}\frac{G^2{\mathcal M}}{c^8}  \int_0^\infty  d\omega\,  \omega^7 \kappa(\omega)
\,,\nonumber\\
(\Delta E)_\mathrm{tail(tail)}&=& 
-\frac{8}{5}\frac{G^3{\mathcal M}^2}{c^{11}}  \int_0^\infty \frac{d\omega}{2\pi}\,\omega^8 \kappa(\omega)\left[   
\ln^2 (C_{I_2} \omega  e^\gamma)-\frac{\pi^2}{12}  +\frac{107}{105}\ln (C_{I_2}  \omega  e^\gamma)+\frac{515063}{176400}
\right]
\,,\nonumber\\
(\Delta E)_\mathrm{(tail)^2}  &=&
\frac{8}{5}\frac{G^3{\mathcal M}^2}{c^{11}} \int_0^\infty \frac{d\omega}{2\pi}\,\omega^{8} \kappa(\omega)\left[
\ln^2(C_{I_2} \omega e^\gamma)+\frac{\pi^2}{4}
\right]
\,,
\eea
and
\bea
\label{final_j_int}
(\Delta J_i)_{\rm tail}&=&
\frac{2}{5}\frac{G{\mathcal M}^2}{c^8} \int_0^\infty d\omega\, \omega^6 \kappa_i(\omega)
\,,\nonumber\\
(\Delta J_i)_{\rm tail(tail)}&=&
-\frac{8}{5}\frac{G^3{\mathcal M}^2}{c^{11}} \int_0^\infty \frac{d\omega}{2\pi}\, \omega^7 \kappa_i(\omega)\left[
\ln^2 (C_{I_2} \omega  e^\gamma)-\frac{\pi^2}{12}  +\frac{107}{105}\ln (C_{I_2}  \omega  e^\gamma)+\frac{515063}{176400}
\right]
\,,\nonumber\\
(\Delta J_i)_{\rm (tail)^2}&=&
\frac{8}{5}\frac{G^3{\mathcal M}^2}{c^{11}} \int_0^\infty \frac{d\omega}{2\pi}\, \omega^7 \kappa_i(\omega)\left[
\ln^2(\omega C_{I_2}e^\gamma) + \frac{\pi^2}{4}
\right]
\,, 
\eea
\end{widetext}
respectively, where we have introduced the notation
\bea
\label{kappatensdef}
\kappa_{ab}(\omega)&=&\hat I_{aj}(\omega) \hat I_{bj} (-\omega)=\kappa_{ba}(-\omega)
\,,\nonumber\\ 
\kappa(\omega)&=&{\rm Tr}[\kappa_{ab}(\omega)]
\,,\nonumber\\
\kappa_i(\omega)&=&2i \epsilon_{iab} \kappa_{ab}(\omega)
\,.
\eea
It is also useful to introduce the magnitude $\mathcal N(\omega)$ and the direction $n_i$ of the vector $\kappa_i(\omega)$, so that
\beq
\label{kappavecdef}
\kappa_i(\omega)\equiv{\mathcal N}(\omega) n_i\,.
\eeq

Notice that: 1)  in both cases the contributions from logarithms squared cancel out once the tail-of-tail and tail-square terms are summed up; for example,
\bea
&&(\Delta E)_\mathrm{tail(tail)+(tail)^2}= 
+\frac{8}{5}\left(-\frac{107}{105}\right)\frac{G^3{\mathcal M}^2}{c^{11}} \times \nonumber\\ 
&&\quad\times \int_0^\infty \frac{d\omega}{2\pi}\,\omega^8 \kappa(\omega) 
\ln  \left( \frac{\omega}{\rm scale}\right)\,,
\eea
where 
\beq
\ln \left(\frac{{\rm scale}}{C_{I_2}e^\gamma}\right)= \frac{105}{107}\left(\frac{\pi^2}{3}-\frac{515065}{176400}\right);
\eeq
 2) the dimensions of $\kappa$ (or $\kappa_i$) are obtained recalling the dimensions of the quadrupolar moment in the Fourier space are not the same as in the ordinary space, namely
\beq
I_{ab}(t)\sim \frac{1}{T}\hat I_{ab}(\omega)\,,
\eeq
with an obvious use of notation.
Therefore
\beq
\kappa(\omega) \sim \hat I^2 \sim (T ML^2)^2\,,
\eeq
which implies for example
\bea
(\Delta E)_{\rm tail} &\sim&  
\frac{G^2M}{c^8} \frac{\kappa(\omega)}{T^8} \nonumber\\
 & \sim & \frac{G^2M}{c^8} \frac{T^2 M^2L^4}{T^8}\nonumber\\
& \sim & Mc^2\,.
\eea

A direct comparison between energy and angular momentum past tails shows that the following simple relation holds between the corresponding densities,
\beq
(\Delta E)_{\rm X} =\int_0^\infty d\omega \frac{dE^{\rm X}}{d\omega}\,,\quad
(\Delta J)_{\rm X} =\int_0^\infty d\omega \frac{dJ^{\rm X}}{d\omega}\,,
\eeq
such that 
\bea
\omega \kappa(\omega)\frac{dJ_i^{\rm X}}{d\omega}-\kappa_i(\omega) \frac{dE^{\rm X}}{d\omega}=0\,,
\eea
for all different tail terms, $X=$ tail, tail(tail) and (tail)$^2$. 
For example,
\beq
\frac{dE^{\rm tail}}{d\omega}=\frac{2}{5}\frac{G^2{\mathcal M}}{c^8} \, \omega^7 \kappa(\omega)\,,
\eeq
etc.
More precisely,
\bea
\label{relEJ}
\frac{dJ_i^{\rm X}}{d\omega}= {\mathcal P}(\omega) \frac{dE^{\rm X}}{d\omega}n_i
\,,
\eea
with
\beq
\label{calPdef}
{\mathcal P}(\omega)\equiv\frac{{\mathcal N}(\omega) }{ \omega \kappa(\omega)}\,,
\eeq
determining both direction and magnitude of the angular momentum flow (in terms of energy flow) in the Fourier space. 
The loss of angular momentum rate per unit frequency thus dominates with respect to the energy one in the range of frequencies wherein  ${\mathcal P}(\omega)>1$, and viceversa for ${\mathcal P}(\omega)<1$.

Equation \eqref{relEJ} connecting the loss of energy and angular momentum rates per unit frequency closely resembles the proportionality relation between the gravitational-wave energy and angular momentum fluxes for circular orbits, satisfying the first law of binary black hole dynamics in the adiabatic approximation \cite{Blanchet:2017rcn}.

\section{Explicit results for hyperboliclike orbits}

Let us evaluate the leading order contribution to the orbital average of the tail integrals \eqref{LOtailsaver} in the case of hyperboliclike motion.
We only need the Newtonian description of the dynamics of a binary system.
The corresponding  Keplerian parametrization of the hyperbolic motion in harmonic coordinates in terms of dimensionless variables (and $c=1$), i.e., $r=r^{\rm phys}/(GM)$, $t= t^{\rm phys}/(GM)$, is \cite{DD1}
\begin{eqnarray} \label{hyporbit}
r&=& \bar a_r (e_r  \cosh v-1)\,,\nonumber\\
\bar n t&=& e_r  \sinh v-v\,,\nonumber\\  
\phi&=&2\, {\rm arctan}\left[\sqrt{\frac{e_r +1}{e_r -1}}\tanh \frac{v}{2}  \right]\,.
\end{eqnarray}
We will assume the motion to be confined in the $x$-$y$ plane, so that $(r,\phi)$ are polar coordinates on that plane.
The expressions of the orbital parameters $\bar n$, $\bar a_r$ and  $e_r$ as functions of the specific binding energy $\bar E$, Eq. \eqref{defbarE}, and of the dimensionless  angular momentum $j$, Eq. \eqref{defj2}, of the system are given by
\beq
\bar n=(2\bar E)^{3/2}\,,\quad
\bar a_r=\frac1{2\bar E}\,,\quad
e_r=\sqrt{1+2\bar Ej^2}\,,
\eeq
with $\bar E>0$ also expressed in terms of the relative momentum for infinite separation $p_\infty$ as $2\bar E\equiv p_\infty^2$.
The parametric equations \eqref{hyporbit} are obtained through analytic continuation of the corresponding elliptic motion ($\bar E<0$) by replacing $v\to iv$.
Therefore, $\bar n$ and $\bar a_r$ are the hyperbolic counterparts of the inverse radial period and the semimajor axis, respectively, whereas $e_r$ still has the meaning of an eccentricity parameter.
Notice that this property is lost from 2PN on \cite{Cho:2018upo}.

The first step consists in Fourier transforming the quadrupole moment, i.e.,
\beq
\label{I_ab_omega2}
\hat I_{ab}(\omega)=\int \frac{dt}{dv}e^{i\omega t(v)}I_{ab}(t)|_{t=t(v)} \, dv \,.
\eeq
This is done by using the integral representation of the Hankel functions of the first kind of order $p \equiv \frac{q}{e_r}$ and argument $q \equiv i \, u$, with 
$u\equiv \omega e_r \bar a_r^{3/2}$, 
\beq
\label{Hankel_rep}
H_p^{(1)}(q)=\frac{1}{ i\pi }\int_{-\infty}^\infty e^{q\sinh v -p v}dv\,.
\eeq
As the argument $q=iu$  of the Hankel function is purely imaginary, the Hankel function becomes converted into a modified Bessel function of the first kind (Bessel $K$ function), according to the  relation
\beq
H_p^{(1)}(iu)=\frac{2}{\pi}e^{-i \frac{\pi}{2}(p+1)}K_p(u)\,.
\eeq
The typical term is of the kind $e^{q\sinh v -(p+k) v}$, the Fourier transform of which is
\beq
e^{q\sinh v -(p+k) v}\to2e^{-i \frac{\pi}{2}(p+k)}K_{p+k}(u)\,,
\eeq
involving Bessel functions having the same argument $u$, but various orders differing by integers.
However,  standard identities valid for Bessel functions allow  one to reduce the orders to either $p$ or $p+1$. 

All energy and angular momentum tail integrals \eqref{final_en_int} and \eqref {final_j_int} are defined in terms of the trace $\kappa(\omega)$ of the tensor $\kappa_{ab}(\omega)$, Eq. \eqref{kappatensdef}, and the magnitude ${\mathcal N}(\omega)$ of its associated vector $\kappa_i(\omega)$ (proportional to its dual and orthogonal to the orbital plane, i.e., with $n_i=\delta_{iz}$), Eq. \eqref{kappavecdef}, which are given by
\begin{widetext}
\bea
\label{kappa_and_calN}
\kappa(u)&=&
32\frac{\nu^2\bar a_r^7}{p^4u^4}e^{-i\pi p}\left\{
u^2(p^2+u^2+1)(p^2+u^2)K_{p+1}^2(u)\right.\nonumber\\
&&
-u\left[(2p-3)u^2+2p(p-1)^2\right](p^2+u^2)K_{p}(u)K_{p+1}(u)\nonumber\\
&&\left.
+\left[u^6+\left(4p^2-3p+\frac13\right)u^4+\left(5p^2-7p+2\right)p^2u^2+2p^4(p-1)^2\right]K_{p}^2(u)
\right\}
\,,\nonumber\\
{\mathcal N}(u)&=&
128\frac{\nu^2\bar a_r^7}{p^4u^4}e^{-i\pi p}
\sqrt{p^2+u^2}\left[uK_{p+1}(u) + (p^2 + u^2 - p)K_{p}(u)\right]\nonumber\\
&&\times
\left\{
(p^2 + u^2)uK_{p+1}(u) - \left[\left(p - \frac12\right)u^2+ p^2(p-1)\right]K_{p}(u)
\right\}
\,,
\eea
\end{widetext}
as functions of the frequency-related variable $u$ introduced above.
For convenience, with an abuse of notation, we denoted here as $\kappa(u)$ the dimensionless (rescaled) quantity
\beq
\kappa(u)\to \frac{\kappa^{\rm phys}(u)}{\left[\left(\frac{GM}{c^2}\right)^3 \frac{M}{c}\right]^2}\,,
\eeq
and similarly for ${\mathcal N}(u)$.

\begin{figure*}
\[
\begin{array}{cc}
\includegraphics[scale=0.35]{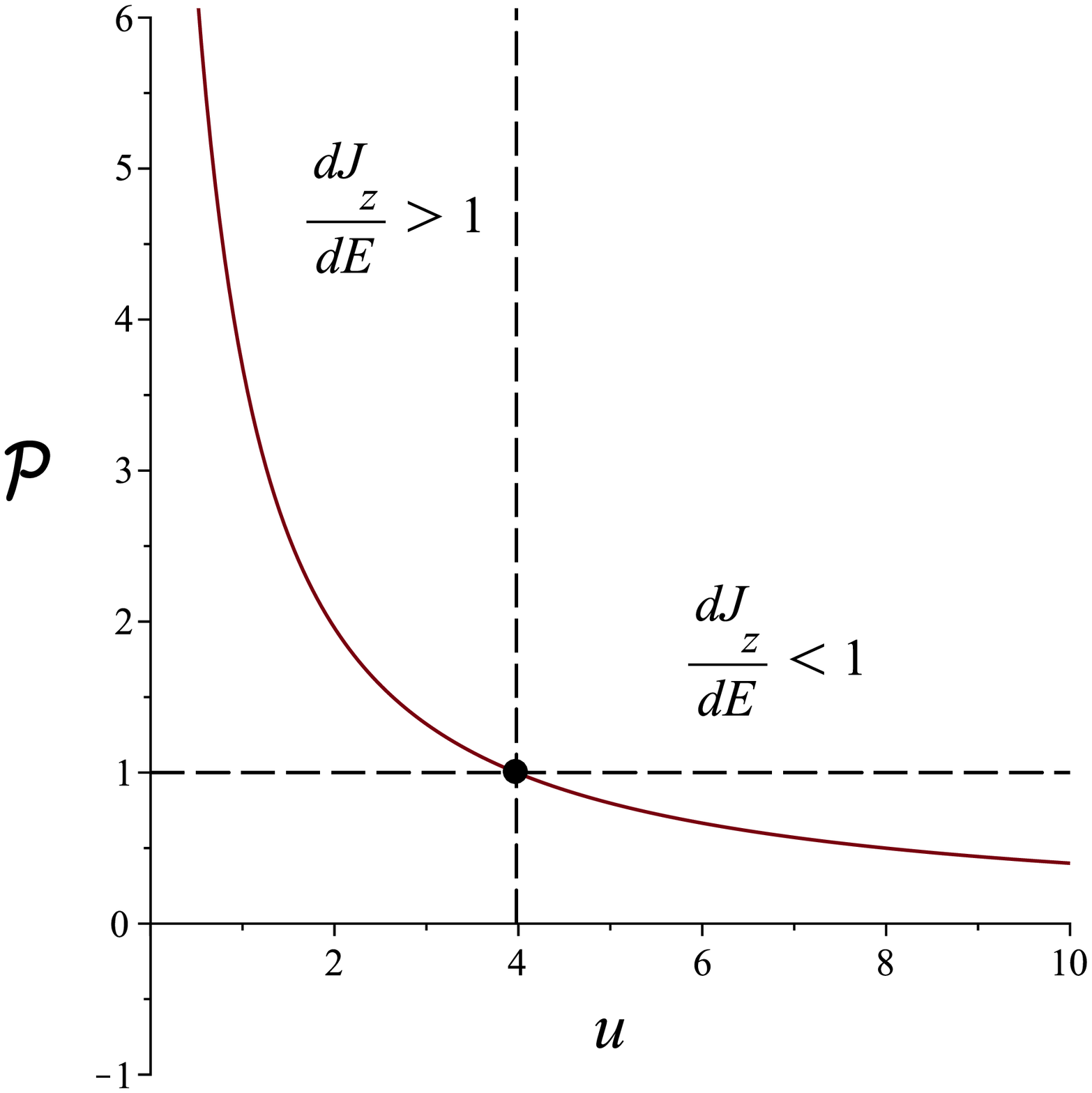}\qquad &  \includegraphics[scale=0.35]{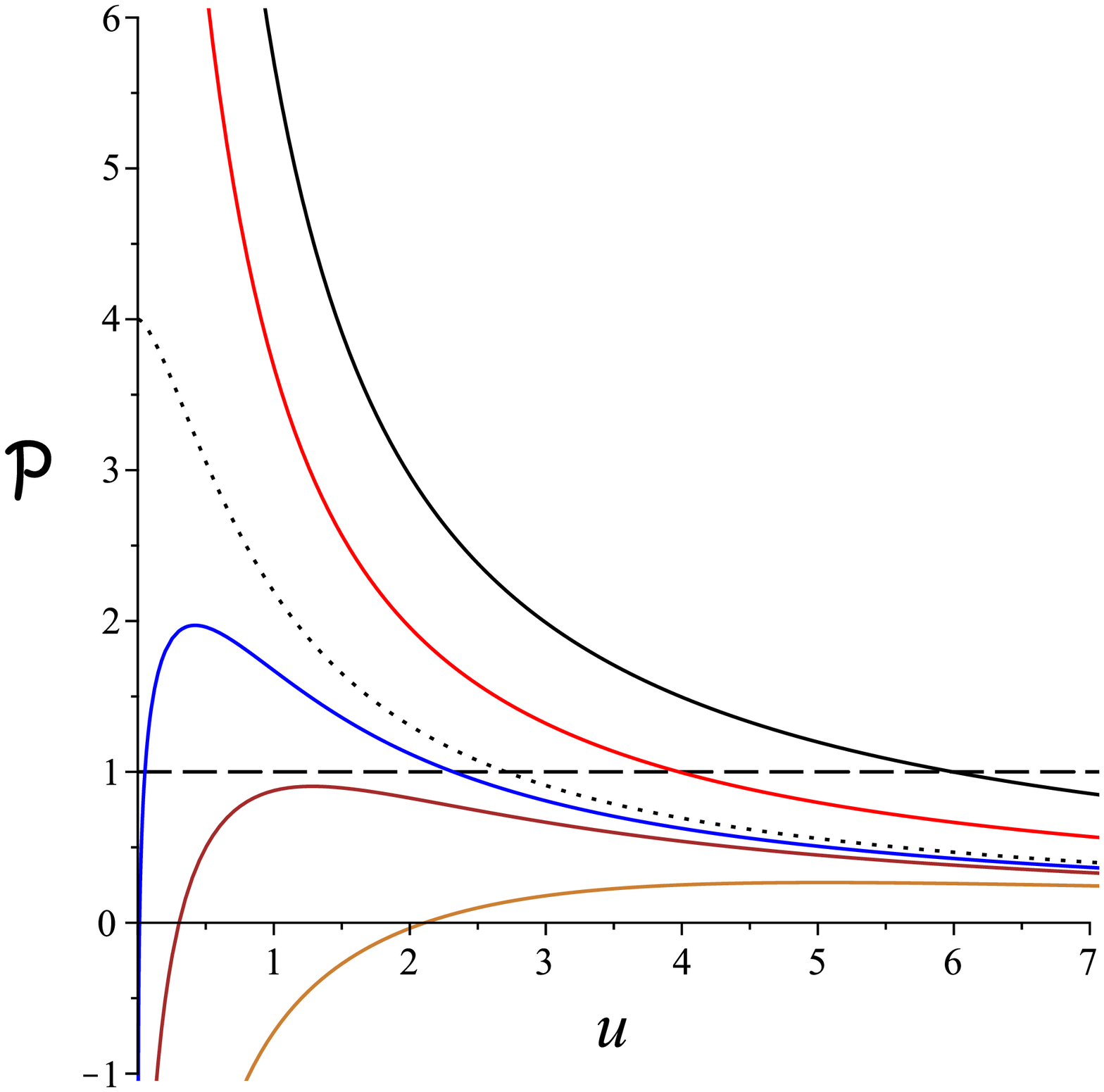}\cr
(a) & (b) 
\end{array}
\]
\caption{\label{fig:1} 
Behavior of the proportionality factor ${\mathcal P}$ (see Eq. \eqref{calPdef}) as a function of the frequency-related dimensionless variable $u$, showing the regions of energy versus angular momentum dominance.
In panel (a) the orbital parameters have been set as $\bar a_r=1$ and $e_r=2$, implying that ${\mathcal P}(u)=1$ at $u=u_*(e_r,\bar a_r)\approx3.979$.
This is the typical behavior for $e_r>e_r^{\rm sep}=\sqrt{2}$, for which ${\mathcal P}$ positively diverges as $u\to0$ and monotonically decreases for increasing frequencies, crossing the horizontal line ${\mathcal P}(u)=1$ at some value of $u=u_*(e_r,\bar a_r)$.
The intersection point moves to the right for increasing values of the eccentricity (see also panel (b)). 
The curves of panel (b) correspond instead to different values of $e_r$ (and the same value of $\bar a_r=1$), moving from bottom to top for increasing eccentricity $e_r=[1.1,1.2,1.3,\sqrt{2},2,3]$.
The dotted curve is the separatrix between the two different behaviors (see text).
}
\end{figure*}

Figure \ref{fig:1} shows the frequency regions of energy versus angular momentum dominance for selected values of the eccentricity parameter and fixed semi-major axis.
At high frequencies ($u\to \infty$) the proportionality factor ${\mathcal P}$ (defined in Eq. \eqref{calPdef}) goes to zero for every value of the eccentricity.
For instance, at the leading order in the large-eccentricity expansion limit we find
\beq
[{\mathcal P}(u)]^{\rm LO}_{u\to\infty}\sim \frac{2}{u}\bar a_r^{3/2} e_r\,,
\eeq
where the asymptotic relation $\bar a_r^{3/2} e_r\sim j/p_\infty^2$ also holds.
In the limit of low frequencies ($u\to 0$) instead its behavior strongly depends on the chosen value of $e_r$.
In fact, for $u\to0$ one has
\bea
{\mathcal P}(u)_{u\to0}&\sim& 2\bar a_r^{3/2}\left[-\frac{e_r^2-2}{\sqrt{e_r^2-1}}\ln\left(\frac{ue^\gamma}{2}\right)\right.\nonumber\\
&&\left. 
+2\sqrt{e_r^2-1}\right]\,,
\eea
implying that there exists a critical value of the eccentricity $e_r=e_r^{\rm sep}=\sqrt{2}$ such that in this limit ${\mathcal P}(u)$ gets the finite value $4\bar a_r^{3/2}$, whereas it logarithmically diverges assuming either positive or negative values depending on whether $e_r$ is greater or smaller than $e_r^{\rm sep}$.


The tail integrals cannot be performed in closed analytical form due to the dependence of the order of the Bessel $K$ functions on the integration variable.
However, the order $p$ tends to zero when $e_r \to \infty$, allowing for the explicit computation in a large-eccentricity expansion \cite{Bini:2017wfr,Bini:2020hmy,Bini:2020rzn}.
In fact, Taylor-expanding the Bessel functions around $p=0$ leads to integrals involving Bessel functions $K_{\nu}(u)$ and their derivatives $\frac{\partial^n K_\nu(u)}{\partial \nu^n }$ with respect to the order $\nu$ evaluated at $\nu=(0,1)$ only. 
Therefore, one is left with integrals of the type
\beq
\int_0^\infty du f(u) \ln^m(u)\,, 
\eeq
with $m=0,1,2$, which are conveniently computed by using the Mellin transform \cite{Bini:2020rzn}.
The latter is defined as 
\beq
\label{mellin}
g(s)=\int_0^\infty du\, u^{s-1} f(u)\,, 
\eeq
so that 
\bea
g(1)&=& \int_0^\infty du f(u) 
\,,\nonumber\\
\frac{dg(s)}{ds}\bigg|_{s=1}&=&\int_0^\infty du f(u)\ln(u)
\,,\nonumber\\
\frac{d^2g(s)}{ds^2}\bigg|_{s=1}&=&\int_0^\infty du f(u)\ln^2(u)
\,.
\eea

We list below the results of the computation, by using the equivalent large-$j$ expansion limit in place of the large-eccentricity limit:
\begin{widetext}
\bea
\label{DE_tail}
(\Delta E)_{\rm tail}&=&Mc^2 
\nu^2\left[\frac{3136}{45} \frac{p_\infty^6}{j^4} 
+  \frac{297\pi^2}{20}\frac{p_\infty^5\pi}{j^5}
+ \left(\frac{9344}{45}  + \frac{88576}{675} \pi^2\right)\frac{p_\infty^4}{j^6} \right.\nonumber\\
&&\left.
+ \left(-\frac{2755}{64} \pi^4 + \frac{1579}{3} \pi^2\right)\frac{p_\infty^3\pi}{j^7} 
+ O\left(\frac{1}{j^8}\right)\right]
\,,\nonumber\\
\label{DE_tail(tail)}
 (\Delta E)_\mathrm{tail(tail)}&=& Mc^2 
\nu^2\left[
\left(-\frac{297}{10}{\mathcal L}^2 - \frac{1709227}{6125} - \frac{130071}{700}{\mathcal L} - \frac{297}{40}\pi^2\right)\frac{p_\infty^8\pi}{j^5}\right.\nonumber\\
&+&\left(- \frac{177152}{225} {\mathcal L}^2   - \frac{841216}{945} {\mathcal L}  
- \frac{2060423552}{826875} -  \frac{1417216}{225}{\mathcal L}\ln(2)   + \frac{44288}{675}\pi^2  - \frac{2834432}{225}\ln(2)^2
\right.\nonumber\\
&&\left.
 - \frac{3364864}{945}\ln(2) \right)\frac{p_\infty^7}{j^6}\nonumber\\
&+&\left(- \frac{405611}{32}\zeta(3) -  \frac{57855}{16}\zeta(3){\mathcal L}  -  \frac{3158}{3}{\mathcal L}^2  - \frac{403863077}{105840} -  \frac{7898921}{2520} {\mathcal L}  -  \frac{1579}{6}\pi^2   +  \frac{8265}{128}\pi^4 \right)\frac{p_\infty^6\pi}{j^7}\nonumber\\
&&
\left.
+O\left(\frac1{j^8}\right)
\right]
\,,\nonumber\\
\label{DE_tail_squared}
(\Delta E)_\mathrm{(tail)^2}&=&Mc^2 
\nu^2\left[
\left(\frac{18093}{160} +\frac{297}{10} {\mathcal L}^2  +  \frac{3111}{20} {\mathcal L}  +  \frac{693}{40}\pi^2\right) \frac{p_\infty^8\pi}{j^5}\right.\nonumber\\
&+& \left( \frac{177152}{225} {\mathcal L}^2 +  \frac{98816}{1125} {\mathcal L}  
+ \frac{7499456}{50625} +  \frac{1417216}{225} {\mathcal L}\ln(2)  +  \frac{44288}{225}\pi^2\right.\nonumber\\
&&\left.   
+  \frac{2834432}{225}\ln(2)^2   + \frac{395264}{1125}\ln(2)  \right)\frac{p_\infty^7}{j^6}\nonumber\\
&+& \left( \frac{173327}{16}\zeta(3)    
 - \frac{55507}{180} 
+  \frac{57855}{16}\zeta(3) {\mathcal L}  +  \frac{3158}{3}{\mathcal L}^2  
+  \frac{82471}{40} {\mathcal L}   +  \frac{11053}{18}\pi^2  -  \frac{8265}{128}\pi^4 \right)\frac{p_\infty^6\pi}{j^7}
\nonumber\\
&&\left.
+O\left(\frac1{j^8}\right)
\right]\,,
\eea
for the energy, and 
 \bea
\label{DJ_tail}
 (\Delta J_z)_{\rm tail}&=& \frac{GM^2}{c} 
\nu^2 \left[\frac{448}{5}  \frac{p_\infty^4}{j^3}
+\frac{69}{5}\pi^2 \frac{p_\infty^3\pi}{j^4}
+\left(\frac{4352}{45}\pi^2+\frac{128}{15} \right)\frac{p_\infty^2}{j^5}
+ \left(-\frac{423}{16} \pi^4 + 303 \pi^2\right)\frac{p_\infty\pi}{j^6} 
+ O\left(\frac{1}{j^7}\right)\right] 
\,,\nonumber\\
\label{DJ_tail(tail)}
 (\Delta J_z)_{\rm tail(tail)}&=&  \frac{GM^2}{c}
\nu^2\left[\left(-\frac{69}{10}\pi^2 - \frac{3997468}{18375} - \frac{27637}{175} {\mathcal L} - \frac{138}{5} {\mathcal L}^2\right)\frac{p_\infty^6\pi}{j^4} \right.\nonumber\\
&&
+\left(-\frac{96745024}{55125} - \frac{8704}{15} {\mathcal L}^2  - \frac{662528}{1575} {\mathcal L}  +  \frac{2176\pi^2}{45} 
-  \frac{139264}{15}\ln(2)^2  -  \frac{2650112}{1575}\ln(2)   - \frac{69632}{15} {\mathcal L} \ln(2) \right)
\frac{p_\infty^5}{j^5}\nonumber\\
&&\left.
+\left(
- \frac{303}{2}\pi^2  - \frac{5175383}{2940} +  \frac{1269}{32}\pi^4   - \frac{104051}{70} {\mathcal L} - 606 {\mathcal L}^2 -  \frac{8883}{4}\zeta(3) {\mathcal L}   - \frac{296757}{40}\zeta(3) 
\right)
\frac{p_\infty^4\pi}{j^6}
+ O\left(\frac{1}{j^7}\right) \right]
\,,\nonumber\\
\label{DJ_tail_squared}
(\Delta J_z)_{\rm (tail)^2}&=&  
\frac{GM^2}{c} 
\nu^2 \left[
\left(\frac{161}{10}\pi^2 + \frac{2833}{40} + \frac{138}{5} {\mathcal L}^2 + \frac{649}{5}{\mathcal L}\right)\frac{p_\infty^6\pi }{j^4}\right.\nonumber\\
&& +\left( \frac{8704}{15} {\mathcal L}^2   -  \frac{512}{3} {\mathcal L}  + \frac{19936}{135} +  \frac{69632}{15} {\mathcal L}\ln(2)   +  \frac{2176}{15}\pi^2   +  \frac{139264}{15}\ln(2)^2  -  \frac{2048}{3}\ln(2) \right)
\frac{p_\infty^5 }{j^5}\nonumber\\
&&\left.+
\left( \frac{707}{2}\pi^2   - \frac{18073}{40} -  \frac{1269}{32}\pi^4   +  \frac{8883}{4}\zeta(3) {\mathcal L}   + 606 {\mathcal L}^2 +  \frac{8689}{10} {\mathcal L}   +  \frac{31437}{5}\zeta(3)  \right)
\frac{p_\infty^4\pi }{j^6}
+ O\left(\frac{1}{j^7}\right)\right]
\,,
\eea
\end{widetext}
for the angular momentum, with
\beq
{\mathcal L}=\ln \left(\frac{r_0 p_\infty^2}{4j}\right)\,.
\eeq

The angular momentum memory integral \eqref{Gimemory} requires a separate treatment.
Let us denote by
\bea
\label{Fbjdef}
F_{bj}(t)&=&\int_{-\infty}^{t} d \tau I_{c b}^{{(3)}}(\tau)\, I_{j c}^{{(3)}}(\tau)\nonumber\\
&=&\int_0^\infty d t' I_{c b}^{{(3)}}(t-t')\, I_{j c}^{{(3)}}(t-t')
\,,
\eea
so that 
\beq
\label{Gimemorynew}
{\cal G}_i^{\rm memory}(t)=
\frac{4}{35}\frac{G^2}{c^{10}}\,\epsilon_{iab}\,I_{aj}^{{(3)}}(t)F_{bj}(t)
\,.
\eeq
The integral \eqref{Fbjdef} does not depend on time. In fact, inserting the Fourier transform of the quadrupole moment yields 
\bea
F_{bj}(t)&=&\int_0^\infty d t' \int_{-\infty}^\infty \frac{d\omega}{2\pi}\int_{-\infty}^\infty \frac{d\omega'}{2\pi}(-i\omega)^3(-i\omega')^3\nonumber\\
&&\times\,  
e^{-i\omega(t-t')}e^{-i\omega'(t-t')}\hat I_{cb}(\omega)\hat I_{jc}(\omega')\nonumber\\
&=&  \int_{-\infty}^\infty \frac{d\omega}{2\pi}\int_{-\infty}^\infty \frac{d\omega'}{2\pi}(-i\omega)^3(-i\omega')^3\nonumber\\
&&\times\,
e^{-i(\omega+\omega') t }\hat I_{cb}(\omega)\hat I_{jc}(\omega')\pi \delta (\omega+\omega')\nonumber\\
&=&  \frac12 
\int_{-\infty}^\infty \frac{d\omega}{2\pi}  \omega^6 \hat I_{cb}(\omega)\hat I_{jc}(-\omega)
\,.
\eea
Recalling the definition of the tensor $\kappa_{ab}(\omega)$, Eq. \eqref{kappatensdef}, and restricting the range of frequencies between $[0,\infty)$ then gives
\beq
F_{bj}=\int_0^\infty \frac{d\omega}{2\pi}  \omega^6 \kappa_{(bj)}(\omega)\,,
\eeq
with $\kappa_{(bj)}(\omega)=\frac12 (\kappa_{bj}(\omega)+\kappa_{jb}(\omega))$\,.
Finally, the orbital average \eqref{LOtailsaver} reads 
\beq
\label{Gimemory2}
(\Delta J_i)_{\rm memory}=
\frac{4}{35}\frac{G^2}{c^{10}} \,\epsilon_{iab}\, H_{aj}\,F_{bj}\,,
\eeq
where 
\beq
H_{aj}=\int_{-\infty}^\infty dt I_{aj}^{{(3)}}(t)\,.
\eeq
The latter integral turns out to be
\bea
H_{aj}&=&
\int_{-\infty}^\infty dt \int_{-\infty}^\infty \frac{d\omega}{2\pi}(-i\omega)^3e^{-i\omega t}\hat I_{aj}(\omega)\nonumber\\
&=& \int_{-\infty}^\infty d\omega(-i\omega)^3 \delta(\omega)\hat I_{aj}(\omega)\nonumber\\
&=&-\frac{4\nu\sqrt{e_r^2-1}}{\bar a_r e_r^2}\, (\delta_{ax}\delta_{jy}+\delta_{ay}\delta_{jx})
\,,
\eea
so that the averaged memory integral \eqref{Gimemory2} becomes
\beq
\label{Gimemory3}
(\Delta J_i)_{\rm memory}=
-\frac{4\nu\sqrt{e_r^2-1}}{\bar a_r e_r^2}(\epsilon_{ixy}F_{yy}+\epsilon_{iyx}F_{xx})\,,
\eeq
with only nonvanishing component
\beq
(\Delta J_z)_{\rm memory}=-\frac{4}{35}\frac{G^2}{c^{10}}\frac{4\nu\sqrt{e_r^2-1}}{\bar a_r e_r^2}(F_{yy}-F_{xx})\,,
\eeq
the large-$j$ expansion of which reads
\bea
(\Delta J_z)_{\rm memory}&=&-\frac{GM^2}{c}\nu^3\left[ \frac{16}{105}\frac{p_\infty^5\pi}{j^4} + \frac{128}{63}\frac{p_\infty^4}{j^5}\right.\nonumber\\
&&\left.
 + \frac{8}{7}\frac{p_\infty^3\pi}{j^6}+O\left(\frac{1}{j^7}\right)\right]\,.
\eea

\section{Time-symmetric tails}

The tails defined above should be more properly termed \lq\lq past tails," since they refer to the past interaction between the two bodies, in the sense that the integration variable $\xi=t-\tau$ in the typical tail integral \eqref{typical_int_n} takes values in the interval $\tau\in[0,\infty)$, namely
\beq
{\mathcal T}_{\ln^m}[X^{(n)}_L;C_{X_L}](t)
=\int_{0}^\infty d\tau X^{(n)}_L(t-\tau)\ln^m \left(\frac{\tau}{C_{X_L}}\right)\,, 
\eeq
implying contributions from $X^{(n)}_L(\xi)$ with $\xi$ varying in the range $\xi\in (-\infty, t]$.

Previous works focusing on ellipticlike motion used the tails in this precise form.
The meaning and importance of time-symmetric tails was proven at the 4PN level in Ref. \cite{Damour:2014jta}, where tail effects on the dynamics were
decomposed in a time-symmetric action contribution and a time-antisymmetric radiation-reaction force.
Such a decomposition seems clearly extendable when considering effects which are linear in radiation-reaction, as recently accomplished in Ref. \cite{Bini:2021gat}. 
It is well known that the radiation-reaction force starts at 2.5PN, so that quadratic effects in radiation-reaction start affecting the dynamics of the system beyond the 4PN order. More precisely, one expects that second-order effects will enter the dynamics at order $\frac{G^4}{c^{10}}$, i.e., at the 4PM level and the 5PN level (see the discussion in Section X of Ref. \cite{Bini:2021gat}).
No complete treatment of the energy flux (as well as angular and linear momentum fluxes) exists at such a level yet.
It is reasonable to expect that the contribution of higher-order time-symmetric tails becomes relevant as soon as the PN accuracy increases, as they did at 4PN. 
Further investigation is necessary to systematically include in the dynamics time-symmetric tails which are nonlinear in radiation-reaction, whatever approach one uses (e.g., the effective field theory approach \cite{Foffa:2011np,Galley:2015kus,Foffa:2019eeb}).

Let us replace $X^{(n)}_L(t-\tau)$ by the sum of its symmetric (sym) and antisymmetric (asym) parts,
\bea
X^{(n)}_L(t-\tau)&=&\frac12[X^{(n)}_L(t-\tau)+X^{(n)}_L(t+\tau)]\nonumber\\
&+&\frac12[X^{(n)}_L(t-\tau)-X^{(n)}_L(t+\tau)]\nonumber\\
&\equiv &  X^{(n)}_{L,\rm sym}(t,\tau) +  X^{(n)}_{L,\rm asym}(t,\tau) \,.
\eea
The time-symmetric part only is used as proper tail contribution, since the time-antisymmetric one is already included in the nonlocal part of the Hamiltonian.
The time-symmetric (ts) version of ${\mathcal T}_{\ln^m}[X^{(n)}_L;C_{X_L}](t)$ thus reads
\bea
&&{\mathcal T}^{\rm ts}_{\ln^m}[X^{(n)}_L;C_{X_L}](t) 
= \nonumber\\
&&\qquad\qquad =\int_{0}^\infty d\tau X^{(n)}_{L\, \rm sym}(t,\tau)\ln^m \left(\frac{\tau}{C_{X_L}}\right)\,. \nonumber\\
\eea  
Passing then to the Fourier domain the above expression becomes
\bea
&&{\mathcal T}^{\rm ts}_{\ln^m}[X^{(n)}_L;C_{X_L}](t) 
= \frac12 \int_{-\infty}^\infty \frac{d\omega}{2\pi}(-i\omega)^n\hat X_L(\omega) \nonumber\\
&&\qquad\times\,
e^{-i\omega  t  }[A_m (\omega, C_{X_L})+A_m (-\omega, C_{X_L})]\,.
\eea
The final expressions for the (averaged) energy and angular momentum time-symmetric tails are given by Eq. \eqref{LOtailsaver} with the replacement ${\mathcal T}_{\ln^m}\to{\mathcal T}^{\rm ts}_{\ln^m}$ in the fluxes \eqref{Ftail_new}--\eqref{Gitail_new}.

Consider now the time symmetric version of the basic integral \eqref{basic_int}, i.e.,
\begin{widetext}
\bea
\label{basic_int_sym}
F_m^{\rm ts}[Y^{(p)}_M,X^{(n)}_L; C_{X_L}]&=&\int_{-\infty}^\infty dt\, Y^{(p)}_M(t)\, {\mathcal T}^{\rm ts}_{\ln^m}[X^{(n)}_L; C_{X_L}](t)\nonumber\\
&=&\int_{-\infty}^\infty dt\, \int_{-\infty}^\infty\frac{d\omega'}{2\pi}e^{-i\omega' t}(-i\omega')^p \hat Y_M(\omega')\nonumber\\
&\times&
\frac12 \int_{-\infty}^\infty \frac{d\omega}{2\pi}(-i\omega)^n\hat X_L(\omega)e^{-i\omega  t  } [A_m (\omega, C_{X_L})+A_m (-\omega, C_{X_L})] \nonumber\\
&=& \frac12 \int_{-\infty}^\infty \frac{d\omega}{2\pi} (-i\omega)^n (i\omega)^p \hat Y_M(-\omega)\hat X_L(\omega)[A_m (\omega, C_{X_L})+A_m (-\omega, C_{X_L})]\,,
\eea
which for $m=1$ and $m=2$ becomes
\bea
\label{basic_int_sym_1}
F_1^{\rm ts}[Y^{(p)}_M,X^{(n)}_L; C_{X_L}]
&=& \frac12 \int_{-\infty}^\infty \frac{d\omega}{2\pi} (-i\omega)^n (i\omega)^p \hat Y_M(-\omega)\hat X_L(\omega)[A_1 (\omega, C_{X_L})+A_1 (-\omega, C_{X_L})]\nonumber\\
&=& \frac12 \int_{-\infty}^\infty \frac{d\omega}{2\pi} (-i\omega)^n (i\omega)^p \hat Y_M(-\omega)\hat X_L(\omega)(-)\frac{\pi}{|\omega|}
\,,
\eea
and 
\bea
\label{basic_int_sym_2}
F_2^{\rm ts}[Y^{(p)}_M,X^{(n)}_L; C_{X_L}]
&=& \frac12 \int_{-\infty}^\infty \frac{d\omega}{2\pi} (-i\omega)^n (i\omega)^p \hat Y_M(-\omega)\hat X_L(\omega)[A_2 (\omega, C_{X_L})+A_2 (-\omega, C_{X_L})]\nonumber\\
&=& \frac12 \int_{-\infty}^\infty \frac{d\omega}{2\pi} (-i\omega)^n (i\omega)^p \hat Y_M(-\omega)\hat X_L(\omega)\frac{2\pi}{|\omega|}\ln (C_{X_L}|\omega| e^\gamma)
\,,
\eea
respectively, having used Eq. \eqref{prop_of_A_n}.
In the special case $Y=X$ and $L=M$ Eqs. \eqref{basic_int_sym_1} and \eqref{basic_int_sym_2} simplify as
\bea
\label{basic_int_sym_1bis}
F_1^{\rm ts}[X^{(p)}_L,X^{(n)}_L; C_{X_L}]
&=& \frac12 (-1)^{n+1} i^{n+p} \int_{-\infty}^\infty \frac{d\omega}{2}  \omega^{n+p}   \hat X_L(-\omega)\hat X_L(\omega) \frac{1}{|\omega|}
\,,\\
\label{basic_int_sym_2bis}
F_2^{\rm ts}[X^{(p)}_L,X^{(n)}_L; C_{X_L}]
&=& \frac12 (-1)^{n} i^{n+p} \int_{-\infty}^\infty  d\omega   \omega^{n+p} \hat X_L(-\omega)\hat X_L(\omega)\frac{1}{|\omega|}\ln (C_{X_L}|\omega| e^\gamma)
\,.
\eea
Therefore, when $n+p$ is odd both  $F_1^{\rm ts}[X^{(p)}_L,X^{(n)}_L; C_{X_L}]$ and $F_2^{\rm ts}[X^{(p)}_L,X^{(n)}_L; C_{X_L}]$ vanish identically.

The time-symmetric energy tail turns out to be 
\bea
\label{final_en_int_ts1}
(\Delta E)_{\rm tail,\,ts}=\int_{-\infty}^\infty dt {\mathcal F}_{\rm tail,\,ts}(t) &=& \frac45 \frac{G^2{\mathcal M}}{c^8}F_1^{\rm ts}[I^{(3)}_{ij},I^{(5)}_{ij}; C_{I_2}]\nonumber\\
&=& \frac45 \frac{G^2{\mathcal M}}{c^8}\frac12 \int_{-\infty}^\infty \frac{d\omega}{2\pi} (-i\omega)^5 (i\omega)^3 \hat I_{ij}(-\omega)\hat I_{ij}(\omega)(-)\frac{\pi}{|\omega|}\nonumber\\
&=& \frac15 \frac{G^2{\mathcal M}}{c^8}  \int_{-\infty}^\infty  d\omega   \omega ^8 \hat I_{ij}(-\omega)\hat I_{ij}(\omega) \frac{1}{|\omega|}\nonumber\\
&=& \frac25 \frac{G^2{\mathcal M}}{c^8}  \int_{0}^\infty  d\omega   \omega ^7 \hat I_{ij}(-\omega)\hat I_{ij}(\omega) \,,
\eea
coinciding with the analogous result for past tails.
The time-symmetric part of the tail-of-tail integral is instead identically vanishing
\bea
\label{final_en_int_ts2}
(\Delta E)_{\rm tail(tail),\,ts}=\int_{-\infty}^\infty dt {\mathcal F}_{\rm tail(tail),\,ts}(t) &=& \frac45 \frac{G^3{\mathcal M}^2}{c^{11}}\left(F_2^{\rm ts}[I^{(3)}_{ij},I^{(6)}_{ij}; C_{I_2}]  -\frac{107}{105} F_1^{\rm ts}[I^{(3)}_{ij},I^{(6)}_{ij}; \tilde C_{I_2}]\right)\nonumber\\
&=& 0\,,
\eea
due to the general property shown above with $n+p=9$. Finally, the time-symmetric tail-squared integral reads 
\bea
\label{final_en_int_ts3}
(\Delta E)_{\rm  (tail)^2,\,ts}&=&\int_{-\infty}^\infty dt {\mathcal F}_{\rm (tail)^2,\,ts}(t) \nonumber\\
&=& \frac15 \frac{G^3{\mathcal M}^2}{c^{11}}\int_{-\infty}^\infty dt  \int_{-\infty}^\infty \frac{d\omega}{2\pi}(-i\omega)^5\hat I_{ij}(\omega)e^{-i\omega t}(-)\frac{\pi}{|\omega|}
\int_{-\infty}^\infty \frac{d\omega'}{2\pi}(-i\omega')^5\hat I_{ij}(\omega')e^{-i\omega' t}(-)\frac{\pi}{|\omega'|}
 \nonumber\\
&=& \frac15 \frac{G^3{\mathcal M}^2}{c^{11}} \int_{-\infty}^\infty \frac{d\omega}{2\pi} (-i\omega)^5(i\omega)^5 \hat I_{ij}(\omega )\hat I_{ij}(-\omega)
\frac{\pi^2}{ \omega^2}\nonumber\\
&=&  \frac15 \frac{G^3{\mathcal M}^2}{c^{11}}\pi \int_0^\infty  d\omega  \omega^8 \hat I_{ij}(\omega )\hat I_{ij}(-\omega)\,,
\eea
differently from the analogous past tail case.

The time-symmetric angular momentum tails turn out to be
\bea
\label{final_j_int_ts}
(\Delta J_i)_{\rm tail,\, ts}=\int_{-\infty}^\infty dt {\mathcal G}_i^{\rm tail,\, ts} &=& \frac{4}{5}\,\frac{G^2 \mathcal{M}}{c^8}\epsilon_{iab}\left[F_1^{\rm ts}[I^{(2)}_{aj},I^{(5)}_{bj};C_{I_2}] +
F_1^{\rm ts}[I^{(3)}_{bj},I^{(4)}_{aj};C_{I_2}] 
 \right]\nonumber\\
&=&  \frac{2}{5}\,\frac{G^2 \mathcal{M}}{c^8}  \int_{0}^\infty  d\omega   \omega^6 \kappa_i(\omega)
\,,\nonumber\\
(\Delta J_i)_{\rm tail(tail),\, ts}=\int_{-\infty}^\infty dt {\mathcal G}_i^{\rm tail(tail),\, ts} &=& \frac{4}{5}\,\frac{G^3\mathcal{M}^2}{c^{11}} \epsilon_{iab}
\left[
F_2^{\rm ts}[I^{(2)}_{aj},I^{(6)}_{bj};C_{I_2}]
+F_2^{\rm ts}[I^{(3)}_{aj},I^{(5)}_{bj};C_{I_2}]\right.\nonumber\\
&&\left.
-\frac{107}{105}\left(F_1^{\rm ts}[I^{(2)}_{aj},I^{(6)}_{bj};\widetilde C_{I_2}]+F_1^{\rm ts}[I^{(3)}_{aj},I^{(5)}_{bj};\widetilde C_{I_2}]\right)
\right] \nonumber\\
&=& 0
\,,\nonumber\\
(\Delta J_i)_{\rm (tail)^2,\, ts}=\int_{-\infty}^\infty dt {\mathcal G}_i^{\rm (tail)^2,\, ts} &=& \frac{8}{5}\,\,\frac{G^3\mathcal{M}^2}{c^{11}}\epsilon_{iab}
\int_{-\infty}^\infty dt \left[\frac12 \int_{-\infty}^\infty \frac{d\omega}{2\pi}(-i\omega)^4\hat I_{aj}(\omega) e^{-i\omega  t  }(-)\frac{\pi}{|\omega|} \right] \nonumber\\
&\times&
\left[\frac12 \int_{-\infty}^\infty \frac{d\omega'}{2\pi}(-i\omega')^5\hat I_{bj}(\omega') e^{-i\omega '  t  }(-)\frac{\pi}{|\omega'|}\right]\nonumber\\
&=&  \frac{\pi }{10}\,\,\frac{G^3\mathcal{M}^2}{c^{11}}   \int_{-\infty}^\infty  d\omega  \omega^7  \kappa_i(\omega)\nonumber\\
&=& 
\frac{\pi }{5}\,\,\frac{G^3\mathcal{M}^2}{c^{11}}   \int_{0}^\infty  d\omega  \omega^7  \kappa_i(\omega) 
\,,
\eea
where we have used the property that $\kappa_i(\omega)$ is an odd function of the frequency (see Eq. \eqref{kappatensdef})
\beq
\kappa_i(-\omega)=2i\epsilon_{iab}\hat I_{aj}(-\omega)\hat I_{bj}(\omega)
=-2i\epsilon_{iba}\hat I_{bj}(-\omega)\hat I_{aj}(\omega)=-\kappa_i(\omega)\,.
\eeq

Summarizing, direct comparison between the orbital averages of the energy and angular momentum tail integrals, Eqs. \eqref{final_en_int} and \eqref{final_j_int}, and their time-symmetric counterparts, Eqs. \eqref{final_en_int_ts1}--\eqref{final_en_int_ts3} and \eqref{final_j_int_ts}, shows that 
\bea
(\Delta E)_{\rm tail,\,ts}&=&(\Delta E)_{\rm tail}\,, \qquad 
(\Delta E)_{\rm tail(tail),\,ts}=0\,, \qquad
(\Delta E)_{\rm  (tail)^2,\,ts}=(\Delta E)_{\rm  (tail)^2}^{\rm nolog}
\,,\nonumber\\
(\Delta J_i)_{\rm tail,\,ts}&=&(\Delta J_i)_{\rm tail}\,, \qquad 
(\Delta J_i)_{\rm tail(tail),\,ts}=0\,, \qquad
(\Delta J_i)_{\rm  (tail)^2,\,ts}=(\Delta J_i)_{\rm  (tail)^2}^{\rm nolog}
\,,
\eea
where \lq\lq nolog'' stands for the nonlogarithmic term of the corresponding (past tail) quantity.
The property \eqref{relEJ} and related discussion apply also to this case.

The explicit computation of the time-symmetric tail-squared integrals in the large-$j$ expansion limit gives
\bea
\label{DE_tail_squared_ts}
(\Delta E)_\mathrm{(tail)^2,\,ts}&=&Mc^2 
\nu^2\pi^2\left[
\frac{297}{40}\frac{p_\infty^8\pi }{j^5}+\frac{44288}{225}\frac{p_\infty^7}{j^6}+\frac{1579}{6}\frac{p_\infty^6\pi }{j^7}
+O\left(\frac1{j^8}\right)\right]\,,
\eea
and 
 \bea
\label{DJ_tail_squared_ts}
(\Delta J_z)_{\rm (tail)^2,\,ts}&=&  
\frac{GM^2}{c} 
\nu^2\pi^2 \left[
\frac{69}{10}\frac{p_\infty^6\pi }{j^4}+\frac{2176}{15}\frac{p_\infty^5 }{j^5}+\frac{303}{2}\frac{p_\infty^4\pi }{j^6}
+ O\left(\frac{1}{j^7}\right)\right]
\,.
\eea
\end{widetext}
Noticeably, the coefficients of the previous expansions,  Eqs. \eqref{DE_tail_squared_ts} and \eqref{DJ_tail_squared_ts},  coincide with the corresponding coefficients of the ${\mathcal L}^2$ terms in Eqs. \eqref{DE_tail_squared} and  \eqref{DJ_tail_squared} divided by a factor of 4.

\section{Concluding remarks}

We have computed higher-order tail (i.e., tail-of-tail and tail-squared) contributions to both the energy and angular momentum losses averaged along hyperboliclike orbits at their leading PN approximation, using harmonic coordinates and working in the Fourier domain.
These terms are conveniently denoted as \lq\lq past tails," since they are determined by the full past interaction among the bodies. 
We have also evaluated the time-symmetric counterpart of these tail integrals, which plays a key role in the construction of the nonlocal part of the conservative two-body dynamics starting from the 4PN level.
All results have been expressed as an expansion in the large-eccentricity parameter, then converted in a large angular momentum expansion.
It is interesting to note that we have obtained a nonvanishing value for the (averaged) nonlinear angular momentum memory integral, differently from the bound case \cite{Arun:2009mc}.

We have also found the interesting result (valid at the same approximation level in which the tail integrals are computed) that there exists a direct proportionality between the loss of energy and angular momentum rates per unit frequency by a frequency-dependent factor which is the same for tails of any kind, generalizing similar links known for circular orbits only. The ranges of frequencies wherein such a factor is smaller/greater than one correspond to those regions in the spectrum of energy/angular momentum loss dominance for fixed values of the orbital parameters. 

The inclusion of tail effects is necessary for the construction of the two-body Hamiltonian at high PN orders as well as for the evaluation of more and more accurate expressions for the radiative losses of energy, angular momentum and linear momentum.
The latter are essential for computing the radiation-reaction contribution to the scattering angle in the relativistic two-body problem \cite{Bini:2021gat}.
We leave for a forthcoming study the computation of higher-order tails of the linear momentum flux, following the same lines outlined in the present work.

\section*{Acknowledgments}
The authors thank  T. Damour for useful discussions.
DB thanks the International Center for Relativistic Astrophysics (ICRA) and the International Center for Relativistic Astrophysics Network (ICRANet) for partial support.
The authors thank Maplesoft$^{\rm TM}$ for providing a complimentary license of MAPLE 2020.
We are indebted to G. Cho for drawing our attention to missing terms in the last Eq. \eqref{DE_tail_squared}.

\end{document}